\documentclass[11pt]{article}
\usepackage[T1]{fontenc}
\usepackage[latin1]{inputenc}
\usepackage[dvips]{epsfig}
\usepackage{graphicx}
\usepackage[english]{babel}
\usepackage{amsmath}
\usepackage{amssymb}
\usepackage{amsfonts}
\title{\bf $f(T)$  quantum cosmology }
\author{F. Darabi\thanks{Email: f.darabi@azaruniv.ac.ir} \; and  K. Atazadeh  \thanks{Email: atazadeh@azaruniv.ac.ir}\\
{\small Department of Physics, Azarbaijan Shahid Madani University, Tabriz, 53714-161 Iran.}}
\usepackage{color}

\begin{document}

\maketitle

\begin{abstract}
We quantize a flat cosmological model in the context of  $f(T)$ theory of modified gravity using the Dirac's quantization approach for Hamiltonian constraint systems. In this regard, first we obtain the Wheeler-DeWitt equation as the operator equation of the Hamiltonian constraint and solve it for some typical cosmological models of $f(T)=T-2\Lambda$, $f(T)= \beta\sqrt{-2T}$
and $f(T)= \gamma T^2$. Then,  in the context of classical-quantum correspondence, we interpret the obtained wavefunctions of the universe to describe an accelerating de Sitter universe which is found to be in good agreement with  $f(T)=T-2\Lambda$
model. Finally, we study Bohm--de Broglie interpretation of the quantum model for  $f(T)=T-2\Lambda$ model.
\end{abstract}
\vspace{2cm}
\section{Introduction}
The current  problems in  standard cosmology, such as  dark energy,  accelerated expansion of the universe, inflation paradigm, and some other related problems have led the people to  introduce and develop the modified theories of gravity. There are many ways to develop modified theories of gravity. The simplest way  is the modification of  Einstein-Hilbert action or corresponding Lagrangian  by  arbitrary functions of the scalars that live on the spacetime manifold. One such modification is the well-known $f(R)$ modified theory of gravity which includes an arbitrary function of the Ricci scalar $R$ \cite{1,2,3,4,5}. Another one is the ``Teleparallel Equivalent of General Relativity'' (TEGR),  so--called $f(T)$ modified theory of gravity,  which includes an arbitrary function of the torsion scalar $T$ \cite{Linder,Myrzakulov,Ferraro,Wu,Iorio}.
The main dynamical variable of TEGR theory or $f(T)$ gravity  is the tetrad or vierbein field, a field of orthonormal basis in the tangent space. The Lagrangian is quadratic in the torsion of the Weitzenb\"{o}ck connection, which is a curvatureless connection that defines a
spacetime with absolute parallelism \cite{Weitzenb}. Because  the action of $f(T)$ gravity includes only first derivatives of the vierbein, the dynamical equations are always second order. Thus, at  field equations level, $f(T)$ gravity is different from $f(R)$ gravity which contains dynamical equations of fourth order.
$f(T)$ theories of gravity have been considered in various cosmological scenarios  in which they can both describe an inflationary expansion without resorting to an inflaton field and  produce an accelerated expansion at late times \cite{Ferraroo}, \cite{Ben}.

Apart from the cosmological interests in the study of $f(T)$ theories of gravity, some interests have been directed toward the study of $f(T)$  gravity in the context of covariant Hamiltonian formalisms to search the degrees of freedom in this modified gravity theory \cite{Ferraro2} and investigate its constraint structure, using the well known Dirac formalism of Hamiltonian constraint system \cite{Dirac,S2F,DHCS}. This line of investigation is particularly relevant to the present paper, as explained below.

The problem of initial conditions is one of the most challenging questions in cosmological models.
Unlike ordinary classical systems in which the dynamical equations are solved by
implementing some initial
conditions, in the case of cosmological models there are no initial conditions external to the universe to
be implemented for solving the Einstein equations. This is mainly because there is no time parameter
external to the universe. This problem may be solved by resorting to the
``quantum cosmology'' where the classical Einstein equations
are replaced by a quantum Schr\"{o}dinger-like equation, so-called Wheeler-DeWitt
equation, subject to some appropriate boundary conditions \cite{QC}.
Quantum cosmology has been studied in the context of various modified gravity theories such as $f(R)$ gravity~\cite{vakili}, massive gravity ~\cite{nima},  rainbow gravity ~\cite{Barun},  conformally coupled scalar field gravity ~\cite{pedram},  Ho\v{r}ava gravity \cite{kord} and so on (see Refs.\cite{darabi}).

To the authors knowledge, the quantum cosmology of $f(T)$  gravity has not
been yet received any attention, so we are motivated in this paper to study the quantum cosmology
of $f(T)$  gravity.  It is well known that the study of  quantum cosmology,
namely the Wheeler-DeWitt equation,  is tightly related to the Dirac formalism of Hamiltonian constraint system \textcolor[rgb]{1,0,0.501961}{\cite{Banerjee}}. This is because of ``{\it Time reparametrization invariance}''
property of  gravitational and cosmological models which make them to be
Hamiltonian constraint systems. Therefore, if we intend to study the quantum cosmology of $f(T)$  gravity, we necessarily need to  implement the Dirac formalism of Hamiltonian constraint system on this modified gravity. However, since
we are merely concerned  about the cosmological variables of $f(T)$  gravity over
a fixed cosmological Friedmann-Robertson-Walker (FRW) background, the implementation
of Dirac formalism on this $f(T)$  cosmological model is  straightforward
and we may start Dirac formalism from beginning, without engaging in the complications of \cite{Ferraro2}.


The outline of this paper is as follows. In section 2, we study the theoretical framework of $f(T )$
gravity and  review the  ``common''  formulation of the model in FRW cosmological background. In section 3, we comment on a subtle inconsistency of this  ``common''  formulation,
and  implement  the Dirac formalism of Hamiltonian constraint systems as
a correct formulation of this modified gravity. In section 4, we quantize the model, according to Dirac, by applying the operator equation of Hamiltonian constraint so-called Wheeler-DeWitt equation, and solve it for some typical cosmological models of
$f(T)$ gravity to obtain the corresponding wavefunction of the Universe. Moreover, in this section by using de-Broglie Bohm interpretation of quantum mechanics, we write the Hamiltonian equations in presence of quantum potential.  The paper ends with a brief conclusion in section 5.

\section{$f(T)$ gravity }
To study of the teleparallel gravity, we use the orthonormal tetrad components
$e_a (x^{\mu})$, where an index $``a$'' runs over $0, 1, 2, 3$ to the
tangent space at each point $x^{\mu}$ of the manifold.
Thus, the relation of the metric $g_{\mu\nu}$ with tetrad components is given by
\begin{equation}
g_{\mu\nu}=\eta_{_{ab}} e^a_\mu e^b_\nu\,,
\label{2.1}
\end{equation}
where $\mu$ and $\nu$ are coordinate indices on the manifold
and also run over $0, 1, 2, 3$, and $e_a^\mu$ forms the tangent vector on the tangent space over which are related to the
metric $\eta_{_{ab}}$.

In the ordinary general relativity we use the torsionless Levi-Civita connection but in the teleparallelism
we use the curvatureless Weitzenb\"{o}ck connection \cite{Weitzenb}, whose non-null torsion $T^\rho_{\verb| |\mu\nu}$ and contorsion
$K^{\mu\nu}_{\verb|  |\rho}$ are given respectively by \cite{TTorsion,TTorsion1,TTorsion2}
\begin{eqnarray}
T^\rho_{\verb| |\mu\nu} \equiv e^\rho_a ~T^a_{\verb| |\mu\nu}= e^\rho_a
\left( \partial_\mu e^a_\nu - \partial_\nu e^a_\mu +\omega^a_{\verb| |b\mu } e^b_\nu-\omega^a_{\verb| |b \nu} e^b_\mu\right)\,,
\label{2.2}
\end{eqnarray}
\begin{eqnarray}
K^{\mu\nu}_{\verb|  |\rho}
\equiv
-\frac{1}{2}
\left(T^{\mu\nu}_{\verb|  |\rho} - T^{\nu \mu}_{\verb|  |\rho} -
T_\rho^{\verb| |\mu\nu}\right)\,,
\label{2.3}
\end{eqnarray}
where $\omega^a_{\verb| |b\mu }$ is the teleparallel {\it spin connection}. 

There are infinite choices of tetrads, so that
one can generally take a particular frame in which all the components of  spin connection vanish. This procedure is  considered as
a different formulation of teleparallel gravity, known as
{\it pure tetrad teleparallel gravity} \cite{TTorsion1,Lucas,Lin,Maluf}. In this formulation,  the tetrad and the
spin connection are considered as  independent variables such that under
a Lorentz transformation the
non-vanishing spin connection can be transformed to vanishing spin connection, independently of transformations of
the tetrad.
Then, the torsion (\ref{2.2}) is redefined in terms of
ordinary derivative, instead of covariant derivative, as follows
\begin{equation}
T^\rho_{\verb| |\mu\nu} \equiv e^\rho_a ~T^a_{\verb| |\mu\nu}= e^\rho_a
\left( \partial_\mu e^a_\nu - \partial_\nu e^a_\mu \right).
\label{2.2.2.2}
\end{equation}
From now on, we will proceed with the formulation of
{\it pure tetrad teleparallel gravity}. Hence, we can define the torsion scalar $T$ as follows
\begin{equation}
T \equiv S_\rho^{\verb| |\mu\nu} T^\rho_{\verb| |\mu\nu}\,,
\label{2.4}
\end{equation}
where
\begin{equation}
S_\rho^{\verb| |\mu\nu} \equiv \frac{1}{2}
\left(K^{\mu\nu}_{\verb|  |\rho}+\delta^\mu_\rho \
T^{\alpha \nu}_{\verb|  |\alpha}-\delta^\nu_\rho \
T^{\alpha \mu}_{\verb|  |\alpha}\right)\,.
\label{2.5}
\end{equation}
Instead of the Ricci scalar $R$ for the Lagrangian density in general relativity, to define the teleparallel Lagrangian density we use the torsion scalar $T$.

Thus, the modified teleparallel $f(T)$ gravity is given by
\begin{equation}
I=\int d^4x |e|  f(T)\,,
\label{2.6}
\end{equation}
where $|e|= \det \left(e^A_\mu \right)=\sqrt{-g}$ and we have put the units as $c=16\pi G=1$. Note that in the action (\ref{2.6}), we have omitted any matter contribution in the action.
Varying of the action~(\ref{2.6}) with respect to the tetrad fields $e_A^\mu$, on can obtain the field equation as~\cite{Bengochea}
\begin{equation}
\frac{1}{e} \partial_\mu \left( eS_A^{\verb| |\mu\nu} \right) f_{T}
-e_A^\lambda T^\rho_{\verb| |\mu \lambda} S_\rho^{\verb| |\nu\mu}
f_{T} +S_A^{\verb| |\mu\nu} \partial_\mu \left(T\right) f_{_{TT}}
+\frac{1}{4} e_A^\nu f = 0\,,
\label{2.7}
\end{equation}
where $f_{_{T}} = \partial f(T) / \partial T$, $f_{_{TT}} = \partial^2 f(T) / \partial T^2$.

In general, in the cosmological study of a torsion-less curvature-full space-time complying the cosmological principles, one can use the {\it closed, flat}
and {\it open} Friedmann-Robertson-Walker (FRW) metric. It has been shown that the curvature-less torsion-full spacetime also allows enough symmetries so that a {\it flat} FRW metric can be studied in this context \cite{FRW}. In this regard, we take the four-dimensional {\it flat} FRW metric as,
\begin{equation}
d s^2 = -N^{2}dt^{2}+a(t)^{2}(dx^{2}+dy^{2}+dz^{2})\,,
\label{2.8}
\end{equation}
where $N$ is the lapse function. In this space-time, $g_{\mu \nu}= \mathrm{diag} (-N^2, a^2, a^2, a^2)$ and
the tetrad components $e^A_\mu = (N,a,a,a)$ yield the exact value of torsion scalar
\begin{equation}\label{torsion}
T=-6\frac{H^2}{N^2},
\end{equation}
where $H=\dot{a}/a$ is the Hubble parameter and the dot denotes for the time derivative.

By choosing $N=1$ in the flat FRW background, from Eq.~(\ref{2.7}) the modified Friedmann equations are given
by~\cite{Bengochea}
\begin{eqnarray}\label{fri}
12f_{_{T}}H^2+f=0\,,
\label{2.9}
\end{eqnarray}
\begin{eqnarray}\label{dot{H}}
\dot{H}=\frac{1}{ 4T\,f_{_{TT}} + 2f_{_{T}} }
\left(-T\,f_{_{T}}+\frac{f}{2} \right)\,.
\label{2.10}
\end{eqnarray}
The first equation is the ``energy constraint'' and the second equation is the
``field equation'' for $H$.
Note that the energy constraint should be imposed on the solutions just after (not before) the field equation is solved for a typical  $f(T)$. It is known that in  $f(T)$ gravity the dynamical equations are always second order. Thus, $f(T)$ gravity is different form the metric $f(R)$  gravity where the gravitational field equation is fourth-order in derivatives. It  seems that the theoretical aspects of $f(T)$ gravity are more interesting than $f(R)$ gravity.

\subsection{Lagrangian formalism}

To consider the $f(T)$ gravity in the FRW background, we can define a canonical point-like Lagrangian ${\cal L}={\cal L}(a,\dot{a}, T,\dot{T})$, where  ${Q}=\{a,T\}$ is the configuration space and ${{\cal T}Q}=\{a,\dot{a}, T, \dot{T}\}$ is the related tangent
bundle on which ${\cal L}$ is defined.  However, since we have the equation (\ref{torsion}) which relates the variable $T$ to the variable $a$, one can use the method of the Lagrange multipliers to set  $T$ as a constraint of the dynamics. By choosing
the suitable Lagrange multiplier and integrating by parts, the Lagrangian ${\cal L}$ becomes canonical. In this model, we have

\begin{equation}\label{action}
I=2\pi^2\int dt\,Na^3 \left\{f(T)-\lambda\left[T+6\left(\frac{\dot a^2}{N^{2}a^2}\right)\right]
\right\} ,
\end{equation}
where $N$ is the lapse function which together with the Lagrange multiplier
$\lambda$, the torsion scalar $T$ and the scale factor $a$ construct the configuration
space  as $\{a, T,  \lambda, N \}$.
The common approach for obtaining the dynamical equations of $f(T)$ gravity
is as follows.
Since the Lapse function is an arbitrary function it is usually fixed  to be $N=1$. Variation with respect to $T$  gives
$\lambda=f_{_{T}}$ which  can be put in the action to yield
\begin{equation}
I=2\pi^2\int dt\,a^3\left\{f-f_{_{T}}\left[T+6\left(\frac{\dot a^2}{a^2}\right)\right]
\right\},
\end{equation}
which is now reduced to the configuration space $\{a, T\}$.  Integrating by parts, gives the point-like FRW Lagrangian
\begin{equation}
{\cal L}= a^3\,(f-f_{_{T}}\,T)
-6\,f_{_{T}}\,a\,\dot a^2
,\label{eqz0}
\end{equation}
which is a canonical function of two coupled fields, $T$ and $a$, both depending on time $t$. The momenta conjugate to variables $a$ and $T$ are
\begin{equation}\label{pa}
p_{_{a}}=\frac{\partial{\cal L}}{\partial\dot{a}}=-12f_{T}a\dot{a}.
\end{equation}
\begin{equation}\label{pT}
p_{_{T}}=\frac{\partial{\cal L}}{\partial\dot{T}}=0.
\end{equation}
The equations of motion for $a$ and $T$ are
obtained  respectively as
\begin{equation}
a^{3}f_{_{TT}}\left(T+6\frac{\dot{a}^2}{a^2}\right)=0,
\end{equation}
\begin{equation}\label{a}
-6f_T\,H^2-12f_T\frac{\ddot{a}}{a}=3(f-f_T\,T)+12f_{_{TT}}\,\dot{T}\,H.
\end{equation}
From the first equation, it turns out that $T$ has no independent dynamics because it is fixed by the dynamics
of $a$ through the constraint
$T=-6{H^2}$.
\subsection{Hamiltonian formalism}
The Hamiltonian can be obtained through Legendre transformation as
\begin{equation}\label{H}
{\cal H}=-\frac{p_{_{a}}^{2}}{24af_{_{T}}}-a^{3}(f-f_{_{T}}T).
\end{equation}
The Hamilton equations are given by
\begin{equation}\label{1}
\dot{a}=\{a,{\cal H}\}=-\frac{p_{_{a}}}{12af_{_{T}}},
\end{equation}
\begin{equation}\label{2}
\dot{T}=\{T,{\cal H}\}=0,
\end{equation}
\begin{equation}\label{3}
\dot{p}_{_{a}}=\{p_a,{\cal H}\}=\frac{p_{_{a}}^{2}}{24a^{2}f_{_{T}}}+3a^{2}(f-Tf_{_{T}}),
\end{equation}
\begin{equation}\label{4}
\dot{p_{_{T}}}=\{p_{_{T}},{\cal H}\}=f_{_{TT}}\left(\frac{p_{_{a}}^{2}}{24af_{_{T}}^{2}}+a^{3}T\right).
\end{equation}

\section{$f(T)$ gravity as a Hamiltonian constraint system}

In this section, we show that there is a subtle inconsistency between  Lagrangian and Hamiltonian formalisms. The inconsistency has its origin in the fact that  the Lagrangian formalism corresponding to the action (\ref{action}) lacks the independent dynamics for the variable $T$ because of the
constraint
$T=-6{H^2}$, whereas the  Hamiltonian formalism corresponding to the action (\ref{action}) determines the independent dynamics $\dot T$ through Eq.(\ref{2}). This inconsistency between Lagrangian and Hamiltonian formalisms
can be resolved by using the Dirac's formalism of Hamiltonian constraint systems \cite{Dirac}.

Our starting point is the  action (\ref{action})
with the Lagrangian
\begin{equation}\label{action'}
{\cal L}=\,Na^3 \left\{f(T)-\lambda\left[T+6\left(\frac{\dot a^2}{N^{2}a^2}\right)\right]
\right\},
\end{equation}
where the configuration
space  is reconsidered as $\{a, T,  \lambda, N \}$, with
$T$, $\lambda$ and $N$ being unfixed as well as $a$.
The conjugate momenta are obtained as
\begin{equation}\label{pa''}
p_{_{a}}=\frac{\partial{{\cal L}}}{\partial\dot{a}}=-\frac{12\lambda a\dot{a}}{N},
\end{equation}
\begin{equation}\label{pT''}
p_{_{T}}=\frac{\partial{ {\cal L}}}{\partial\dot{T}}=0,
\end{equation}
\begin{equation}\label{p-lambda}
p_{_{\lambda}}=\frac{\partial{ {\cal L}}}{\partial\dot{\lambda}}=0,
\end{equation}
\begin{equation}\label{pN}
p_{_{N}}=\frac{\partial{ {\cal L}}}{\partial\dot{N}}=0.
\end{equation}
The Hamiltonian is constructed, using Legendre transformation, as
\begin{equation}
H_0=p_{_{a}} \dot a +p_{_{T}} \dot T +p_{\lambda} \dot \lambda+p_{_{N}}\dot{N}-{\cal L}=-N\left(\frac{p_{_{a}}^{2}}{24a\lambda}+a^{3}(f(T)-\lambda
T)\right),
\end{equation}
which leads to the Hamilton equations  given by

\begin{equation}\label{3-}
\dot{p}_{_{a}}=\{p_a,{H_0}\}=\frac{p_{_{a}}^{2}}{24a^{2}f_{_{T}}}+3a^{2}(f-Tf_{_{T}}),
\end{equation}
\begin{equation}\label{4-}
\dot{p_{_{T}}}=\{p_{_{T}},{H_0}\}=N a^{3} (f_{_{T}}-\lambda),
\end{equation}
\begin{equation}\label{5-}
\dot{p}_{_{\lambda}}=\{p_{_{\lambda}},{H_0}\}=-N\left(\frac{p_{_{a}}^{2}}{24a\lambda^2}+a^{3}T\right),
\end{equation}
\begin{equation}\label{6-}
\dot{p_{_{N}}}=\{p_{_{N}},{H_0}\}=\frac{p_{_{a}}^{2}}{24a\lambda}+a^{3}(f(T)-\lambda
T),
\end{equation}
\begin{equation}\label{1'}
\dot{a}=\{a,{H_0}\}=-N\frac{p_{_{a}}}{12a\lambda},
\end{equation}
\begin{equation}\label{2'}
\dot{T}=\{T,{H_0}\}=0,
\end{equation}
\begin{equation}\label{3'}
\dot{\lambda}=\{\lambda,{H_0}\}=0,
\end{equation}
\begin{equation}\label{4'}
\dot{N}=\{N,{H_0}\}=0.
\end{equation}
The last three results which determine the dynamics of $\{T, \lambda, N \}$
are still inconsistent with the physical content of the action (\ref{action}) which
apparently lacks any dynamics of $\{T, \lambda, N \}$. Hopefully, the inconsistency is removed provided
that we consider the $f(T)$ gravity as a Hamiltonian constraint system and
generalize the original Hamiltonian $H_0$, by adding some system constraints, according
to Dirac's formalism.

In this regard, we consider the equations (\ref{pT''}), (\ref{p-lambda}) and
(\ref{pN}) as the {\it Primary}
constraints \cite{Dirac}
\begin{equation}\label{pT'}
\phi_{_{T}}=p_{_{T}}\approx 0,
\end{equation}
\begin{equation}\label{p-lambda'}
\phi_{_{\lambda}}=p_{_{\lambda}}\approx0,
\end{equation}
\begin{equation}\label{p-N'}
\phi_{_{N}}=p_{_{N}}\approx0.
\end{equation}
The total Hamiltonian is constructed by adding these primary constraints
to the original Hamiltonian
\begin{equation}\label{Htot}
H_{_{Tot}}=H_0+u^{_{T}}\phi_{_{T}}+u^{\lambda}\phi_{\lambda}+u^{_{N}}\phi_{_{N}},
\end{equation}
where $u^{_{T}}$,  $u^{\lambda}$, and $u^{_{N}}$ are arbitrary coefficients.
Using $H_{_{Tot}}$, the Hamilton equations yield
\begin{equation}\label{3--}
\dot{p}_{_{a}}=\{p_a,H_{_{Tot}}\}=\frac{p_{_{a}}^{2}}{24a^{2}f_{_{T}}}+3a^{2}(f-Tf_{_{T}}),
\end{equation}
\begin{equation}\label{4--}
\dot{p_{_{T}}}=\{p_{_{T}},H_{_{Tot}}\}=N a^{3} (f_{_{T}}-\lambda),
\end{equation}
\begin{equation}\label{5--}
\dot{p}_{_{\lambda}}=\{p_{_{\lambda}},H_{_{Tot}}\}=-N\left(\frac{p_{_{a}}^{2}}{24a\lambda^2}+a^{3}T\right),
\end{equation}
\begin{equation}\label{6--}
\dot{p_{_{N}}}=\{p_{_{N}},H_{_{Tot}}\}=\frac{p_{_{a}}^{2}}{24a\lambda}+a^{3}(f(T)-\lambda
T),
\end{equation}
\begin{equation}\label{1''}
\dot{a}=\{a,{H_{_{Tot}}}\}=-N\frac{p_{_{a}}}{12a\lambda},
\end{equation}
\begin{equation}\label{2''}
\dot{T}=\{T,{H_{_{Tot}}}\}=u^{_{T}},
\end{equation}
\begin{equation}\label{3''}
\dot{\lambda}=\{\lambda,{H_{_{Tot}}}\}=u^{\lambda},
\end{equation}
\begin{equation}\label{4''}
\dot{N}=\{N,{H_{_{Tot}}}\}=u^{_{N}}.
\end{equation}
Now, the appearance of arbitrary coefficients $u^{_{T}}$,  $u^{\lambda}$, and $u^{_{N}}$ in the last three equations, compared to (\ref{2'}), (\ref{3'}), and (\ref{4'}), account for the arbitrary dynamics of $\dot{T}$,  $\dot{\lambda}$ and $\dot{N}$,  in complete agreement with the physical content of the action
 (\ref{action}), and hence the above mentioned inconsistency is removed. Note however that the other dynamical equations have not been changed by adding the
constraints to the original Hamiltonian. The powerful formalism of Dirac's Hamiltonian constraint systems is now ready for the full study of $f(T)$ gravity. The consistency conditions
for the primary constraints
as\begin{equation}\label{2'''}
\dot{\phi_{_{T}}}=\{\phi_{_{T}},{H_{_{Tot}}}\}\approx 0,
\end{equation}
\begin{equation}\label{3'''}
\dot{\phi_{\lambda}}=\{\phi_{\lambda},{H_{_{Tot}}}\}\approx 0,
\end{equation}
\begin{equation}\label{4'''}
\dot{\phi_{_{N}}}=\{\phi_{_{N}},{H_{_{Tot}}}\}\approx 0,
\end{equation}
lead to the {\it secondary} constraints as
\begin{equation}\label{2''''}
{\chi_{_{T}}}=Na^3\left( f_{{_T}}-\lambda\right)\approx 0,
\end{equation}
\begin{equation}\label{3''''}
{\chi_{_{\lambda}}}=-N\left(\frac{p_{_{a}}^2}{24
\lambda^2 a}+a^3 T\right)\approx 0,
\end{equation}
\begin{equation}\label{4''''}
{\chi_{_{N}}}=\left(\frac{p_{_{a}}^{2}}{24a\lambda}+a^{3}(f(T)-\lambda
T)\right)\approx 0.
\end{equation}
Both the primary and secondary constraints
can be considered as six constraints $\phi_j\approx0, (j=1,...,6).$ The consistency conditions
for the secondary constraints also lead to the following  equations

 \begin{equation}\label{2'''''}
\dot{{\chi_{_{T}}}}=\{{\chi_{_{T}}},{H_{_{T}}}\}\approx 0\Longrightarrow u^{m}\{{\chi_{_{T}}}, \phi_m\}\approx-\{\chi_{_{T}},{H_{0}}\},
\end{equation}
\begin{equation}\label{3'''''}
\dot{\chi_{_{\lambda}}}=\{\chi_{{_{\lambda}}},{H_{_{T}}}\}\approx 0\Longrightarrow u^{m}\{\chi_{_{\lambda}}, \phi_m\}\approx-\{\chi_{_{\lambda}},{H_{0}}\},
\end{equation}
\begin{equation}\label{4'''''}
\dot{{\chi_{_{N}}}}=\{{\chi_{_{N}}},{H_{_{T}}}\}\approx 0\Longrightarrow u^{m}\{{\chi_{_{N}}}, \phi_m\}\approx-\{\chi_{_{N}},{H_{0}}\},
\end{equation}
or
\begin{equation}\label{2''''''}
a^2(Nu^{_{T}}f_{_{TT}}-Nu^{\lambda}+u^{_{N}}(f_{{_T}}-\lambda))\approx\frac{N^2p_{_{a}}}{4\lambda}
(f_{{_T}}-\lambda),
\end{equation}
\begin{equation}\label{3''''''}
Nu^{_{T}}a^3-Nu^{\lambda}\frac{p_{_{a}}^2}{12\lambda^3
a}+u^{_{N}}\left(\frac{p_{_{a}}^2}{24
\lambda^2 a}+a^3 T\right)\approx\frac{N^2p_{_{a}}}{4\lambda}
(2T-{f}/{{\lambda}}),
\end{equation}
{\begin{equation}\label{4''''''}
u^{_{T}}a^3(f_{{_T}}-\lambda)-u^{_{\lambda}}\left(\frac{p_a^2}{24
\lambda^2 a}+a^3 T\right)\approx0,
\end{equation}
which can be considered as a set of inhomogeneous equations to determine  the
arbitrary coefficients  $u^{_{T}}=U^{_{T}}$,  $u^{\lambda}=U^{\lambda}$ and
$u^{_{N}}=U^{_{N}}$. These solutions are
not unique and one can consider the homogeneous equations
\begin{equation}\label{2'''''''}
 V^{m}\{{\chi_{_{T}}}, \phi_m\}\approx0,
\end{equation}
\begin{equation}\label{3'''''''}
 V^{m}\{\chi_{_{\lambda}}, \phi_m\}\approx0,
\end{equation}
\begin{equation}\label{4'''''''}
 V^{m}\{{\chi_{_{N}}}, \phi_m\}\approx0,
\end{equation}
to find the new independent solutions $V^{m}_a, (a=1,...,A)$.
These solutions can be added through the arbitrary functions $v^a$ to the previous ones to obtain the general solutions $u^m=U^m+v^aV^{m}_a$ \cite{Dirac}.
The determinant of  coefficient matrix $\Delta$ corresponding to the homogeneous
equations is vanishing  and the $\Delta$ matrix becomes singular.
It also turns out that the {\it Rank} and {\it Nullity} of
 $\Delta$ matrix
is equal to 2 and 1, respectively. Considering all these together,  one
finds that there is only one nontrivial solution vector
($a=1$) for the homogeneous equations
as
\begin{equation}
V^{m} \equiv(V^{_{T}}=0,  V^{\lambda}=0, V^{_{N}}= arbitrary),
\end{equation}
which yields
\begin{equation}
u^{m} \equiv(U^{_{T}},  U^{\lambda}, U^{_{N}}+vV^{_{N}}).
\end{equation}
Up to now, we have 3 {\it primary} and 3 {\it secondary} constraints
together with 2 determined and 1 undetermined Lagrange coefficients, respectively
as $(U^{_{T}},  U^{\lambda})$ and
$(U^{_{N}}+vV^{_{N}}$).
It is time to determine which of them are {\it first class} and which of them are {\it second class} constraints.
From the structures of all constraints $\phi_j\approx0$, it turns out that:
\begin{itemize}
\item The {\it primary} constraint  $\phi_{N}\approx 0$ has weakly vanishing
Poisson brackets with all  constraints $\phi_j\approx0\:$$ (j=1,...,6)$. Therefore it is a  {\it first class} constraint.
\item
All other constraints $\phi_{_{T}}\approx 0$, $\phi_{_{\lambda}}\approx 0$,  $\chi_{_{T}}\approx 0$ and $\chi_{_{\lambda}}\approx 0$  are   {\it second class} constraints.
\item The {\it first class} constraint $\phi_{_{N}}\approx 0$ is the generator of gauge dynamics $\delta N=\epsilon^{^N} \{N, \phi_{_{N}}\}=\epsilon^{^N}$, where $\epsilon^{^N}$ is an arbitrary time dependent coefficient.
\end{itemize}

Up to this stage the gauge invariant quantities due to one  {\it first class} constraint $\phi_{_{N}}\approx 0$  are $(a, p_{_a}, \lambda, T)$ because of $\delta a=\epsilon^{^N} \{a, \phi_{_{N}}\}=0$,  $\delta p_{_a}=\epsilon^{^N} \{p_{_a}, \phi_{_{N}}\}=0$,
$\delta \lambda=\epsilon^{^N} \{\lambda, \phi_{_{N}}\}=0$ and $\delta T=\epsilon^{^N} \{T, \phi_{_{N}}\}=0$, respectively. Gauge invariance of  the physical variables $(a, p_{_a})$ is trivial, and hence gauge invariance of $(\lambda, T)$ is justified by $\lambda=f_{_{T}}$,  $T=g(a, p_{_{a}})$, as a result of $\chi_{_{T}}\approx 0$ and $\chi_{_{\lambda}}\approx 0$.
Here, we note that the system of {\it second class}  constraints  $\phi_{_{T}}\approx 0$, $\phi_{_{\lambda}}\approx 0$,  $\chi_{_{T}}\approx 0$ and $\chi_{_{\lambda}}\approx 0$  can be equivalent with  a system of  {\it first class} constraints  $\phi_{_{T}}\approx 0$, $\phi_{_{\lambda}}\approx 0$   where $\chi_{_{T}}$ and $\chi_{_{\lambda}}$ play the roles of gauge variables\footnote{To show how this works,  consider a simple Lagrangian $L=\frac{1}{2}\dot{q}^2_1-q_2^2$
from which one can easily find the physical variables as $(q_1, p_1)$ and the second class constraints ($q_2\approx0$, $p_2\approx0$). Now, one can reconsider the second class
constraint  $q_2\approx0$ as
the ``gauge fixing'' condition for the gauge transformation generated
by the first class constraint $p_2\approx 0$, namely $\delta q_2=\epsilon^{^N} \{ q_2, p_2\}=\epsilon^{^N}$. In other words, if one does not impose
the ``gauge fixing'' condition $q_2\approx 0$, then we have a system of first
class constraint $p_2\approx 0$ ($q_2$ being a gauge variable, $\delta q_2=\epsilon^{^N} $) which is equivalent to the original system of second class constraints ($q_2\approx0$, $p_2\approx0$).
This is because, in the equivalent system of first class constraint we have
the same physical variables $(q_1, p_1)$ as those of present in the  original system of second class constraints.}. Now, altogether in this equivalent system of first class constraints, we have three first class constraints $(\phi_{_{N}}\approx 0, \phi_{_{T}}\approx 0$, $\phi_{_{\lambda}}\approx 0)$. The total Hamiltonian is  obtained by using $u^{m} \equiv(U^{_{T}},  U^{\lambda}, U^{_{N}}+vV^{_{N}})$ and  imposing $\lambda=f_{_{T}}$,  $T=g(a, p_{_{a}})$ on $H_0$ to obtain

\begin{equation}
H_{_{Tot}}=-N\mathcal{H}+U^{_{T}}\phi_{_{T}}+U^{\lambda}\phi_{\lambda}+vV^{_{N}}\phi_{_{N}},
\end{equation}
where
\begin{equation}\label{mathcalH}
\mathcal{H}=\left(\frac{p_{_{a}}^{2}}{24af_{_{T}}
}+a^{3}(f(T)-f_{_{T}}
T)\right)|_{_{T=g(a, p_{_{a}})}}\:.
\end{equation}
In this form, the constraints $\phi_{_{T}}\approx0$ and $\phi_{\lambda}\approx0$
which were considered as {\it second class} constraints are reconsidered as equivalent  {\it first class} constraints, and   $U^{_{T}}$ and $U^{_{\lambda}}$
which were considered as  determined coefficients corresponding to the {\it second class} constraints, are reconsidered as  determined coefficients corresponding to the gauge fixings  $\lambda=f_{_{T}}$ and $T=-6H^2$, respectively. This gauge property of   $U^{_{T}}$ and $U^{_{\lambda}}$ lets us to interpret them equally as undetermined coefficients of the   {\it first class} constraints
 $\phi_{_{T}}\approx0$ and $\phi_{\lambda}\approx0$.

Now, time independence of the {\it first class}  constraint $\phi_{_{N}}$ leads
to
\begin{equation}\label{H-C}
\dot{\phi}_{_{N}}=\{{\phi}_{_{N}}, \bar{H}_{_{Tot}}\}=\mathcal{H}\approx0,
\end{equation}
where $\mathcal{H}\approx0$ is considered as a {\it first class} constraint,
so called ``{\it Hamiltonian constraint}'',
which  involves just the {\it physical variables} $(a, p_{_{a}})$.
The equations of motion for the physical variables are now obtained as
\begin{equation}\label{711}
\dot{a}=\{a, \bar{H}_{_{Tot}}\}=-N\frac{p_{_{a}}}{12af_T},
\end{equation}
\begin{equation}
\dot{p_{_{a}}}=\{p_{_{a}}, H_{_{Tot}}\}=-N\left(\frac{p^{2}_a}{24a^{2}f_T}-3a^{2}(f-Tf_T)\right),
\end{equation}
\begin{equation}\label{2''-}
\dot{T}=\{T,{H_{_{Tot}}}\}=U^{_{T}},
\end{equation}
\begin{equation}\label{3''-}
\dot{\lambda}=\{\lambda,{H_{_{Tot}}}\}=U^{\lambda},
\end{equation}
\begin{equation}\label{4''-}
\dot{N}=\{N,{H_{_{Tot}}}\}=vV^{_{N}}.
\end{equation}
The last three equations
 show that  ${T}$,  ${\lambda}$ and ${N}$ have gauge dynamics, due to the gauge property of   $U^{_{T}}$ and $U^{_{\lambda}}$ and arbitrariness    of $vV^{_{N}}$, in complete agreement with the essence of the  {\it first class}  constraints
 $\phi_{_{T}}\approx0$,  $\phi_{\lambda}\approx0$ and $\phi_{_{N}}\approx0$.
Especially, gauge
dynamics
of the lapse function
accounts for the ``{\it Time Reparametrization Invariance}'' of the gravitational
model.

 \section{Quantization of $f(T)$ cosmology as the Hamiltonian constraint system}

Up to now, we have shown that the only variables which are viable in the
classical study of $f(T)$ cosmology, as a classical Hamiltonian constraint system, are $a$, ${T}$,  ${\lambda}$ and ${N}$. One can then define a {\it Minisuperspace}
$\{a, T, \lambda, N\}$ over which the  $f(T)$ quantum cosmology can be formulated.
The origin of this {\it Minisuperspace} lies in the action (\ref{action}) or the Lagrangian (\ref{action'}).

The  method of quantization is
essentially involves restricting the Hilbert space in the quantum theory to ensure
that constraints are obeyed by the state vectors. This is called ``{\it Dirac quantization of constraint systems}''. State vectors which satisfy this property
are called {\it physical states} and the sector of the original Hilbert space spanned by these physical states is
called the {\it physical state space}.
In our case, such property is satisfied by the operator version of the {\it
first class constraints} as
\begin{equation}\label{hatT}
\hat{\phi_{_{T}}}|\Psi\rangle=0,
\end{equation}
\begin{equation}\label{hatlambda}
\hat{\phi_{_{\lambda}}}|\Psi\rangle=0,
\end{equation}
\begin{equation}\label{hatN}
\hat{\phi_{_{N}}}|\Psi\rangle=0,
\end{equation}
\begin{equation}\label{hatH}
\hat{\mathcal{H}}|\Psi\rangle=0,
\end{equation}
where $|\Psi\rangle$ is the {\it physical state}. The first tree equations,
using  $p_{_{T}}\rightarrow  -i\frac{\partial}{\partial T}, p_{_{\lambda}}\rightarrow  -i\frac{\partial}{\partial \lambda}, p_{_{N}}\rightarrow  -i\frac{\partial}{\partial N}$,  guarantee that $|\Psi\rangle$ is independent of $\{T, \lambda, N \}$
and so it is just a function of the scale factor $a$.
In the context of quantum cosmology,
the operator equation (\ref{hatH})
and the {physical state} $|\Psi\rangle$  are considered as the Wheeler-DeWitt equation and the wavefunction $\Psi(a)=\langle
 a|\Psi\rangle$, respectively,  in the
{\it Minisuperspace}.
The Wheeler-DeWitt equation is specifically resulted by using of $p_{_{a}}\rightarrow  -i\frac{\partial}{\partial a}$ and operator ordering in the the operator equation (\ref{hatH}).
In order to find the explicit form of the Wheeler-DeWitt equation, we need to suggest the appropriate forms of the function $f(T)$ in the classical
Hamiltonian constraint (\ref{H-C}).
There are some candidates for $f(T)$ gravity, from different points of view,
especially as  alternatives to other modified gravity theories like $f(T)$ gravity.
In the following,  we limit ourselves to  some typical forms of $f(T).$

\subsection{$f(T)=T-2\Lambda$}

 We consider this model in agreement with  the  observational considerations  for current accelerating phase of the universe, where a cosmological term $\Lambda$ is responsible for this acceleration. In fact,
this $f(T)$ gravity is equivalent to the general relativity with a cosmological
constant, namely $R-2\Lambda$, which  describes
a de Sitter accelerating universe.
This accelerating phase can  be derived, as well, by putting $f(T)=T-2\Lambda$ in the equation
(\ref{dot{H}})  which results in
\begin{equation}
2\dot{H}-3H^2+\Lambda=0,
\end{equation}
and has a solution $H=\sqrt{\Lambda/3}$ expressing de Sitter expansion.
One of the  interesting topics in  quantum cosmology is the prediction of
 classical limit. In this regard, we shall study the quantum cosmology of
$f(T)=T-2\Lambda$ and try to interpret the corresponding wavefunction of the universe  that can describe an accelerating classical universe.
Putting this  $f(T)$ into the constraint equation (\ref{H-C}) we obtain
\begin{equation}\label{W1}
\frac{\partial^2\Psi(a)}{\partial a^2}+\frac{q}{a}\frac{\partial\Psi(a)}{\partial a}+24 \Lambda a^{4}\Psi(a)=0,
\end{equation}
where $q$ is the operator ordering parameter. The analytic solutions of Eq.(\ref{W1})  can be expressed in terms of the Bessel functions $J$ and $Y$ as follows
\begin{equation}\label{Ws}
\Psi(a)=\left(\frac{2}{3}\Lambda\right)^{\frac{1-q}{12}} a^{\frac{1-q}{2}}\left[c_1J_\frac{1-q}{6}\left(2 \sqrt{\frac{2\Lambda}{3}} a^3\right)+c_2Y_\frac{q-1}{6}\left(2 \sqrt{\frac{2\Lambda}{3}} a^3\right)\right].
\end{equation}
According to \cite{Vilenkin}, its nonsingular boundary is the line $a = 0$, while at the singular boundary this variable is infinite.
Now, we impose the boundary condition on the above solutions such that at  $a=0$  the wave function vanishes to avoid the singularity  \cite{Vilenkin}. This
yields $c_2 = 0$, and  by choosing $c_1 = 1$ we arrive at the unique solution

\begin{equation}\label{Ws2}
\Psi(a)=\left(\frac{2}{3}\Lambda\right)^{\frac{1-q}{12}} a^{\frac{1-q}{2}}J_\frac{1-q}{6}\left(2 \sqrt{\frac{2\Lambda}{3}} a^3\right).
\end{equation}

It is worth mentioning  that Eq.(\ref{W1}) is a Schr\"{o}dinger-like equation which describes the motion of a fictitious particle with zero energy
under the superpotential $U(a) = -24\Lambda a^{4}$. In general, and for a typical
superpotential $U(a)$,
the minisuperspace may be divided into two regions, $U(a) > 0$ and $U(a) < 0$, which can be termed as the
classically forbidden and classically allowed regions, respectively. The classically forbidden region corresponds to the
exponential behavior of the wavefunction,  while in the classically allowed region the wavefunction
has oscillatory behavior. The division of minisuperspace into classically forbidden and classically allowed regions makes it possible that the Universe
can tunnel from ``nothing'' to the ``existence'', similar to the tunneling effect  through a potential barrier in the
sense of usual quantum mechanics \cite{Vilenkin}.

In our model, however, there is no possibility of quantum tunneling  because the superpotential is always negative and there is no a potential barrier  through which the Universe can tunnel
from ``nothing'' to ``existence''.
Therefore, the wave function always exhibits oscillatory behavior to mimic
a classical evolution of the Universe. The appearance of cosmological constant $\Lambda$ both in the amplitude and the
argument of Bessel functions $J$ in the solution (\ref{Ws2}) is of particular importance  which deserves further discussion in the following.

In the figures 1 and 2, we have plotted
the square of  wavefunction (\ref{Ws2}) for the typical values  ($q=-1$, $\Lambda=1$) and  ($q=-1$, $\Lambda=8$), respectively.

\begin{figure*}[ht]
  \centering
  \includegraphics[width=2.5in]{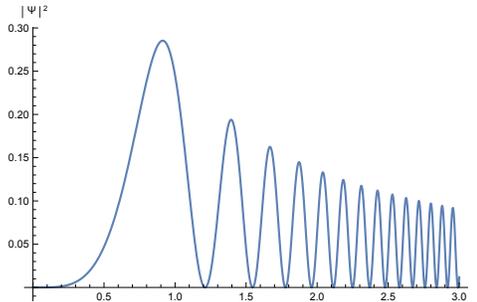}\hspace{2cm}
  \caption{The square of the wave function for the quantum universe with $q=-1$ and $\Lambda=1$.}
  \label{stable1}
\end{figure*}

\begin{figure*}[ht]
  \centering
  \includegraphics[width=2.5in]{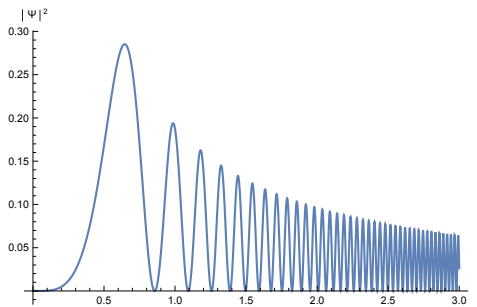}\hspace{2cm}
  \caption{The square of the wave function for the quantum universe with $q=-1$ and $\Lambda=8$.}
  \label{stable2}
\end{figure*}
The following general properties are seen in the figures:
\begin{itemize}
\item
The wavefunction has a well-defined behavior near $a = 0$ and describes a universe, without singularity problem, emerging out of
nothing without any tunneling.
\item For large cosmological constants, the locations of all amplitudes are
shifted towards $a = 0$ and the frequency of oscillation is increased. This property cause the amplitude to more decrease at large scale factors.
\end{itemize}

\subsection{Typical $f(T)$ }

We may consider  other typical forms of  $f(T)$ and study the corresponding
quantum cosmologies. In doing so, we put any desired form of $f(T)$ into the equation (\ref{H-C}) to obtain the explicit form of the Hamiltonian constraint
and then derive the corresponding Wheeler-DeWitt equation to obtain the desired wavefunction. Here, for simplicity, we consider two common forms   $f(T)= \beta\sqrt{-2T}$
and $f(T)= \gamma T^2$ where $\beta$ and $\gamma$ are dimensional constants.
After straightforward calculations, it turns out that for    $f(T)= \beta\sqrt{-2T}$
model the superpotential  vanishes. For $f(T)= \gamma T^2$ model
 the superpotential becomes a nonvanishing function of $T$, however using
$T=-6H^2$ and $\dot{a}=-\frac{p_{_{a}}}{24a\gamma }$  (see (\ref{711}) with
$N=1$),  it can
be rewritten in terms of $p_a^2$.   
Therefore, using  factor ordering, the Wheeler-DeWitt equation for these models leads to
\begin{equation}\label{W11}
\frac{\partial^2\Psi(a)}{\partial a^2}+\frac{q}{a}\frac{\partial\Psi(a)}{\partial a}=0,
\end{equation}
where the  superpotential $U(a)$ is absent. The analytic solutions of Eq.(\ref{W11}) is given by
\begin{equation}\label{Ws222}
\Psi(a)=c_{1}\frac{a^{1-q}}{1-q}+c_{2}.
\end{equation}}
\begin{figure*}[ht]
  \centering
  \includegraphics[width=2.5in]{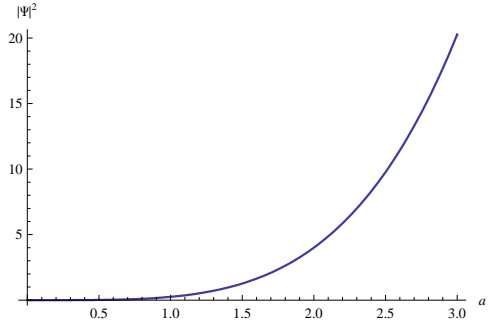}\hspace{2cm}
  \caption{The square of  wave function for the quantum universe. We have put $q=-1$, $c_1=1$ and $c_2=0$.}
  \label{stable}
\end{figure*}
In the figure 3, we have plotted the square of  wavefunction (\ref{Ws222}).

The following general property is seen in this figure:
\begin{itemize}
\item
The wavefunction has  a well-defined behavior near $a = 0$ but fails to be
square-integrable function.
The divergent behavior of wavefunction at $a \rightarrow \infty$ prevents it to be considered as a {\it good}  wavefunction.
\end{itemize}

\subsection{Classical limit}

One of the most challenging topics in  quantum cosmology is the mechanisms
through which the classical cosmology can be predicted by quantum cosmology. Most of the suggestions in resolving this problem use the properties of  wavefunction. In this regard, we try to find the suitable interpretations of the obtained wavefunctions,
using their properties.

\begin{itemize}
\item{$f(T)=T-2\Lambda$}
\end{itemize}
Considering the above mentioned second property of the wavefunction (\ref{Ws2}), we easily
find that this wavefunction describes appropriately a classical universe which tends
to be realized (from nothing)  at smaller scale factors for  larger values
of cosmological constants. In other words, the probability of ``{\it realization from nothing}''
becomes larger for larger values of cosmological constants, in agreement with the results
obtained for the probability of ``{\it tunneling from nothing}'' \cite{LambdaDecay}. This property  coincides with the inflationary scenario in that the universe having a large cosmological constant emerges from nothing with large probability,
at small scales, and this is just the right initial condition for inflation, namely once the universe with large cosmological constant is realized from nothing at small scale, it immediately
begins a de Sitter inflationary expansion. The accelerating behavior of de Sitter expansion is manifested within the ``decreasing amplitude'' and the ``increasing frequency'' of the wavefunction, in terms of the scale factor, in both figures. These  behaviors mimic the accelerating motion of a zero-energy particle under a negative gravitational potential.

\begin{itemize}
\item{Typical $f(T)$}
\end{itemize}

 Unlike the model $f(T)=T-2\Lambda$, the models   $f(T)= \beta\sqrt{-2T}$
and $f(T)= \gamma T^2$ fail to represent a de Sitter inflationary expansion,
through the classical-quantum correspondence, due to  not well-defined wavefunction (\ref{Ws222}).

\subsection{Bohm--de Broglie interpretation of the quantum model}
In the context of the Bohm--de Broglie interpretation of quantum mechanics and also its application in quantum cosmology,
we may use the polar form of the wave function $\Psi(a) =\Omega(a)e^{iS(a)}$ in the corresponding wave equation to obtain the modified Hamilton-Jacobi equation as
\begin{equation}
{\cal H}\left(q_i,p_i=\frac{\partial S}{\partial q_i}\right)+{\cal Q}=0,
\end{equation}
where $p_i$ and ${\cal Q}$ are the momentum conjugate to the dynamical variables $q_i$ and the quantum potential, respectively.
 Following the arguments in
the previous section about the preference of  $f(T)=T-2\Lambda$ model,  here we just focus on this model for which the wave equation (\ref{W1}),  in the
context of Bohm--de Broglie interpretation, can be written as
\begin{equation}
\frac{1}{24a}\left(\frac{\partial S}{\partial a}\right)^{2}+2\Lambda a^{3}-{\cal Q}=0,
\end{equation}
where the quantum potential is defined as
\begin{equation}
{\cal Q}=\frac{1}{24a\Omega}\frac{\partial^{2}\Omega}{\partial a^{2}}+\frac{q}{24a^{2}\Omega}\frac{\partial\Omega}{\partial a}.
\end{equation}
Thus, the quantum Hamiltonian is given by
\begin{equation}
{\cal H}_{{\cal Q}}={\cal H}+{\cal Q},
\end{equation}
where ${\cal H}$ is the gauge fixed Hamiltonian over the reduced phase space
$(a, p_a)$. The quantum equations of motion over the reduced phase space
 are
obtained as\begin{equation}
\dot{a}=\{a, {\cal H}_{{\cal Q}}\}=-\frac{p_{_{a}}}{12a},
\end{equation}
\begin{equation}
\dot{p_{_{a}}}=\{p_{_{a}}, {\cal H}_{{\cal Q}}\}=\frac{p^{2}_a}{24a^{2}}+3\Lambda
a^{2}-\frac{\partial{\cal Q}}{\partial a}.
\end{equation}
From the above equations we obtain
\begin{equation}\label{222}
p_a=-12a\dot{a}\,,
\end{equation}
\begin{eqnarray}\label{dotH}
\dot{H}
=-\frac{1}{  2 }
\left(\frac{T+2\Lambda}{2}+\frac{1}{6a^{2}}\frac{\partial{\cal Q}}{\partial a}\right)\,.
\end{eqnarray}
Putting $T=-6{H^2}$
in the equation
(\ref{dotH})   results in
\begin{equation}\label{2dot{H}}
2\dot{H}-3H^2+(\Lambda+\frac{1}{6a^{2}}\frac{\partial{\cal Q}}{\partial a})=0,
\end{equation}
which shows that the quantum potential can alter the contribution of
cosmological constant in the cosmic dynamics of de Sitter expansion.
Also, the quantum Hamiltonian constraint ${\cal H}_{{\cal Q}}= 0$ leads to
\begin{eqnarray}\label{fri2}
6H^2-2\Lambda+\frac{{\cal Q}}{a^{3}}=0\,.
\end{eqnarray}
Both equations (\ref{2dot{H}}) and (\ref{fri2}) indicate that the contribution of quantum potential to the cosmic dynamics, in the context of Bohm--de Broglie interpretation, is vanishing at large scale factors (when the universe is considered
as a classical system) and
is very important at early universe (when the universe is considered
as a quantum system).

\section{Conclusions}
We have quantized a flat cosmological model in the context of  $f(T)$ theory of
modified gravity. First, we have shown that the correct study of $f(T)$ gravity should
be analyzed using the formalism of  Dirac's Hamiltonian constraint systems. Then, we have proceed to quantize this model using the Dirac's quantization approach for Hamiltonian constraint systems.
We have obtained the Wheeler-DeWitt equations  for  typical cosmological
models of $f(T)=T-2\Lambda$,   $f(T)= \beta\sqrt{-2T}$
and $f(T)= \gamma T^2$, and interpreted the obtained wavefunctions
to find which of them can be preferred to describe an accelerating universe in the context
of classical-quantum correspondence. Finally, we have studied Bohm--de Broglie interpretation of the quantum model for the preferred model $f(T)=T-2\Lambda$.

\section*{Acknowledgment}
This work has been supported by a grant/research fund number 217/D/17738 from Azarbaijan Shahid Madani
University.

\end{document}